\documentstyle[prc,aps,floats,epsfig,preprint]{revtex}

\begin{document}

\title{
           Statistical evolution of isotope composition of nuclear 
           fragments.}
 
\author{
         A.S.~Botvina$^{a,b}$ and I.N.~Mishustin$^{c,d}$}

\address{ 
 $^a$Gesellschaft f\"ur Schwerionenforschung, D-64291 Darmstadt, Germany\\ 
 $^b$Institute for Nuclear Research, 117312 Moscow, Russia\\
 $^c$Niels Bohr Institute, DK-2100 Copenhagen, Denmark\\
 $^d$Kurchatov Institute, Russian Research Center, 123182 Moscow, Russia
}

\date{\today}

\maketitle

\begin{abstract}
Calculations within the statistical multifragmentation model show that 
the neutron content 
of intermediate mass fragments can increase in the region of liquid-gas 
phase transition in finite nuclei. The model predicts also inhomogeneous 
distributions of 
fragments and their isospin in the freeze-out volume caused by an angular 
momentum and external long-range Coulomb field. These effects can take 
place in peripheral 
nucleus-nucleus collisions at intermediate energies and lead to neutron-rich 
isotopes produced in the midrapidity kinematic region.
\end{abstract}

\vspace{1.0cm}
PACS number(s):     
25.70.Pq, 24.10.Pa, 25.70.Mn

\newpage

Multifragmentation of nuclei is the most interesting phenomenon in 
intermediate energy nuclear physics. This is a promising process for studying 
nuclear matter properties at extreme conditions of high excitation energies, 
small densities and at different isospins. In particular, one hopes to 
establish its connection to a nuclear liquid--gas phase transition 
\cite{hirschegg}. As other complicated many-body processes this phenomenon 
can be successfully treated in a statistical way \cite{PR95,gross97}. 
Fragment production in both peripheral \cite{botvina95,dagostino99} 
and central \cite{dagostino96,williams} collisions has clear statistical 
features, though a considerable preequilibrium emission and collective energy 
(e.g. radial flow) should be taken into consideration. 
However, in finite-size nuclear systems statistical processes can lead to 
specific effects since the fragment formation is governed by both short-range 
nuclear forces and long-range Coulomb forces. We show in this paper how the 
interplay of these different forces is manifested 
in isotope production. Also we show that an external Coulomb field, e.g. 
a Coulomb interaction of the target and projectile-like sources 
in peripheral nucleus--nucleus collisions, 
can provide a non-isotropic statistical emission of fragments 
and cause asymmetry in the fragment isospin distribution. 

The statistical multifragmentation model (SMM) is described in detail in many 
publications, see e.g. \cite{PR95}. 
The model is based upon the assumption of statistical equilibrium at a 
low-density freeze-out stage. 
We consider all break-up channels (partitions) composed of nucleons and 
excited fragments taking into 
account mass, charge, momentum and energy conservations. In the microcanonical 
treatment the statistical weight of decay channel j is given by 
$W_{j} \propto exp~S_{j}$, where $S_{j}$ is the entropy of 
the system in channel $j$ depending on excitation energy $E_s^{*}$, 
mass number $A_s$, charge $Z_s$ and other parameters of the source. 
Light fragments with mass number $A\leq 4$ are considered as stable 
particles ("nuclear gas") with only translational degrees of freedom; 
fragments with $A > 4$ are treated as heated nuclear liquid  drops. 
In the standard version \cite{PR95} the Coulomb interaction between 
the fragments is treated in the Wigner--Seitz approximation. 
Different break-up partitions are sampled according to their statistical 
weights uniformly in the phase space. 
After break-up of the nuclear source 
the fragments propagate independently in their mutual Coulomb field and
undergo secondary decays. 
The deexcitation of the hot primary fragments
proceeds via evaporation, fission or Fermi-break-up \cite{botvina87}. 

A new version of SMM \cite{mc_pre,mc_smm} is based on generating the Markov 
chain of partitions which is representative of the whole partition 
ensemble \cite{markov}. 
Individual partitions are generated and selected into the 
chain by applying the Metropolis algorithm and taking into account that 
fragments with the same mass $A$ and charge $Z$ are indistinguishable. 
Due to a high efficiency of this method one can directly calculate the Coulomb 
interaction energy for each spatial configuration of primary 
fragments in the freeze-out volume. 
In this way one can take into account the 
correlations between positions of the primary fragments and their Coulomb 
energy that influences the partition probabilities \cite{gross97}. 
Also one can calculate the moment of inertia and take into account the 
angular momentum conservation similar to Refs.\cite{gross97,bot_gro95}. 
The new method is consistent with the previous one \cite{PR95} 
based on the Wigner--Seitz approximation. The mean parameters of produced 
fragments obtained in the standard and the Markov chain versions 
fit each other reasonably well. 
The full analysis of the Markov chain SMM appears elsewhere \cite{mc_smm}, 
here we present our new results concerning the isospin degree of freedom 
in nuclear multifragmentation 
and provide some guidelines for future studies. 
We concentrate on properties of hot primary fragments since, apart of very 
light fragments, the secondary decay doesn't change the predicted isospin 
evolution.

\begin{figure}
\includegraphics[width=16cm]{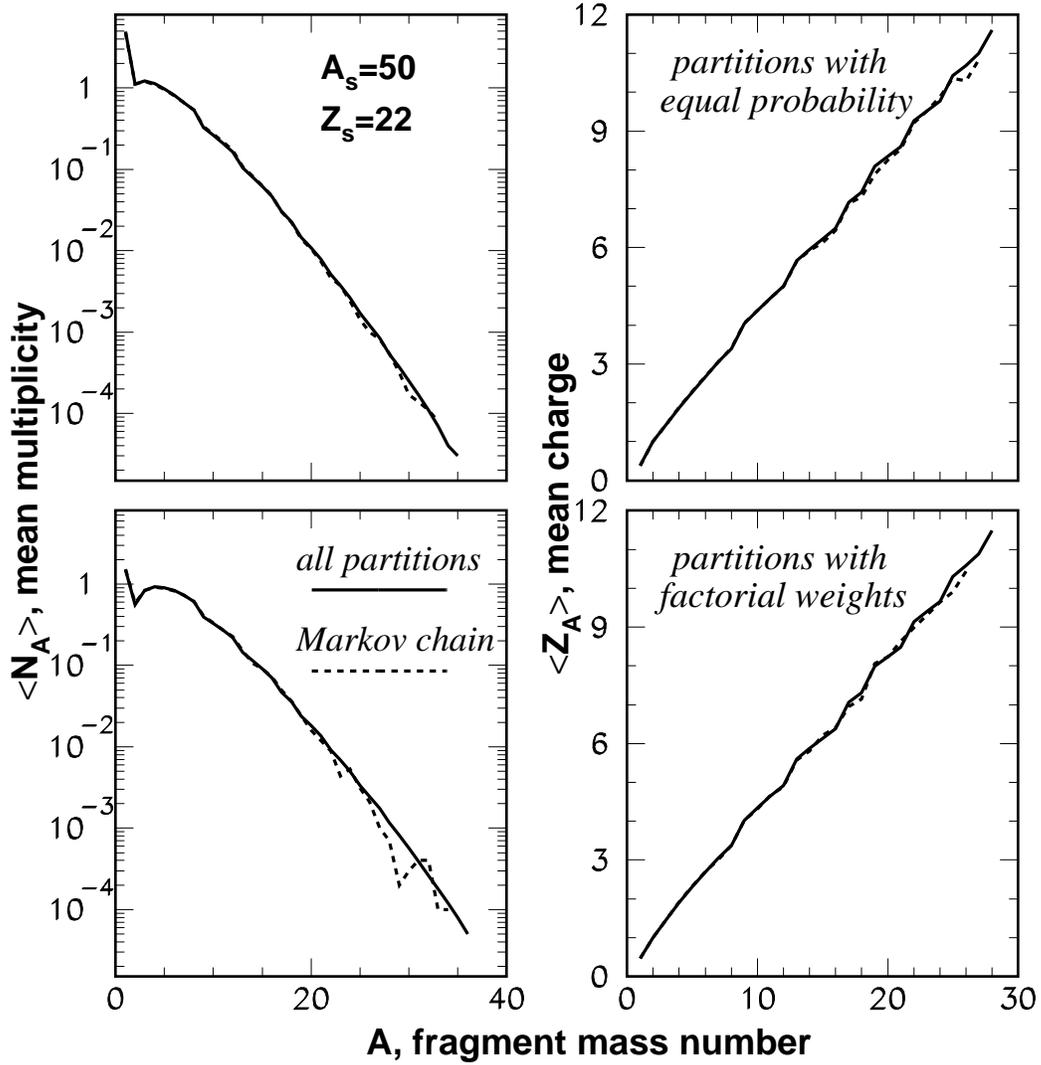}
\caption{Mean multiplicities and charges of fragments obtained by the
Markov chain generation of partitions (dashed lines) as compared with
direct accounting for all partitions (solid lines), for the system with 
total mass number $A_s$=50 and charge $Z_s$=22. In the top panels 
partitions are taken with equal probability, in the bottom ones partitions 
have factorial weights.}
\end{figure}

Before any application it is important to demonstrate that the
Monte--Carlo generation of the Markov chain exactly represents the whole
ensemble of fragment partitions. The case of one component system 
(that takes into account only nucleons without isospin) was considered in 
\cite{mc_pre}. Here we present results for the two-component system when 
fragments are characterized by both the mass number and charge. The best way 
to check the numerical algorithm is to compare it with the direct calculation 
accounting for all possible partitions, one by one. As shown in \cite{mc_pre}, 
the precision of the Monte--Carlo algorithm is very good and does not depend 
significantly on the system's size. Fig.~1 shows the distributions of average 
fragment multiplicities and average charges versus mass number of fragments 
for two cases which differ by the partition weights. 
In the first case all partitions are taken 
with equal weights, while in the second case, with the factorial weights, 
$W=1/\prod N_{AZ}!$, where $N_{AZ}$ is the multiplicity of fragments with mass 
number $A$ and charge $Z$. Despite of the fact that these weights are 
essentially different, the results obtained in the direct and Markov chain 
calculations coincide in both cases. The same good agreement holds for other 
weights and fragment partition characteristics.

In the beginning we consider simple thermal effects concerning isospin 
distribution in multifragmentation. In grand canonical limit the formulae for 
the average charge $\langle Z_{A} \rangle$ of fragment $A$ and the 
width $\sigma^{A}_{Z}$ of the charge distribution were obtained in 
\cite{botvina87}:
\begin{equation}
\langle Z_{A} \rangle =\frac{(4\gamma+\nu)A}{8\gamma +2cA^{2/3}},~~
\sigma^{A}_{Z} \approx \sqrt \frac{AT}{8\gamma},
\end{equation}
where $T$ is the temperature, $\gamma\approx$25 MeV is the symmetry energy 
coefficient, $\nu$ is the chemical potential responsible for the charge 
conservation. A Coulomb energy of the system is accounted in the 
mean-field Wigner-Seitz approximation leading to 
$c=(3/5)(e^2/r_0)(1-(\rho/\rho_0)^{1/3})$, where $e$ is the proton charge, 
$r_0$=1.17 fm, $\rho$ is the nuclear freeze-out density and 
$\rho_0\approx$0.15 fm$^{-3}$ is the normal nuclear density. 
The $A$-dependence of $\langle Z_{A} \rangle$ recalls a famous experimental 
systematics for the stable nuclei \cite{viola}: 
$\langle Z_{stable} \rangle = A/(1.98+0.0155A^{2/3})$. The difference is in 
the additional chemical potential $\nu$ and in the reduced Coulomb 
coefficient. As calculations show, $\nu$ decreases with the 
neutron excess of the source so that the fragments produced from a 
neutron-rich source are also neutron-rich (see e.g. Fig.~3 in \cite{milazzo}). 

\begin{figure}
\includegraphics[width=16cm]{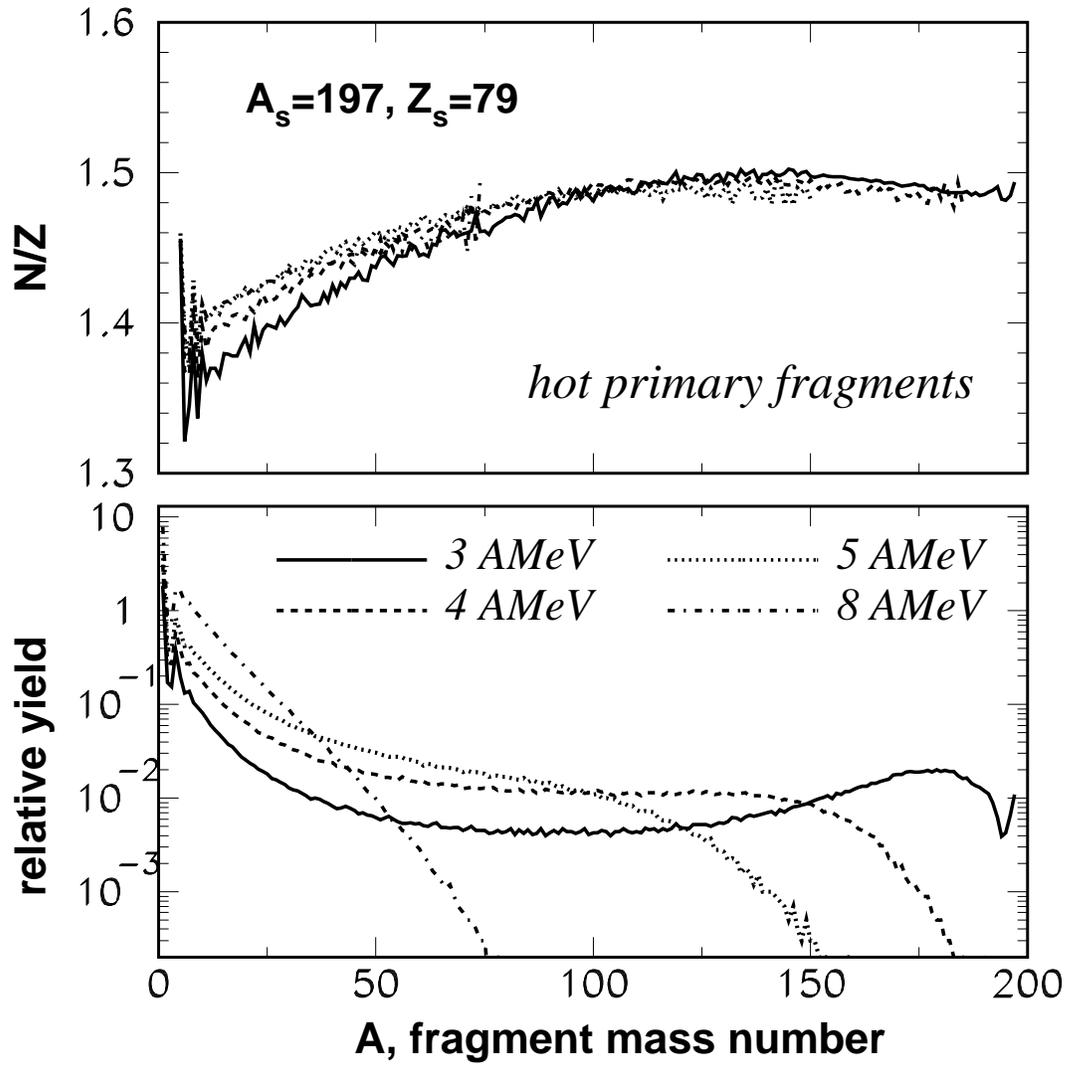}
\caption{The neutron-to-proton ratio N/Z and relative yield of 
hot primary fragments produced after break-up of Au nucleus at 
different excitation energies: 3 (solid lines), 4 (dashed lines), 
5 (dotted lines) and 8 (dot-dashed lines) MeV per nucleon.} 
\end{figure}

In finite systems isospin of the fragments changes with the excitation 
energy in a special way. In Fig.~2 we show neutron-to-proton ratios (N/Z) 
of the fragments together with their mass distributions. 
The calculations were done with the Markov 
chain SMM for Au source ($A_s$=197, $Z_s$=79) at density $\rho_s$=$\rho_0$/3 
using the Wigner-Seitz approximation as in 
previous standard SMM calculations \cite{milazzo}. 
However, the direct evaluation of the Coulomb interaction 
does not change the presented results. One can see a general statistical 
trend: the N/Z ratio of hot primary fragments increases with their mass 
numbers, as is also evident from formula (1). This is a consequence of the 
interplay between 
Coulomb and symmetry energy contributions to the binding energy of 
fragments \cite{PR95}. This trend persists up to $A \leq A_s/2$, at 
larger $A$ the finite-size effects due to the mass and charge 
conservation demolish it. In Fig.~2 
we demonstrate also the evolution of N/Z ratio and mass distribution of 
fragments in the excitation energy range $E_s^*$=3--8 MeV/nucleon. 
It is seen that the fragment mass distribution evolves from the U--shape, 
at the multifragmentation threshold $E_s^* \sim 3$ MeV/nucleon, 
to an exponential fall at high energies. As the energy increases 
the N/Z ratio of primary 
intermediate mass fragments (IMF, charges Z=3--20) increases too. 
The reason is that the heaviest neutron-rich fragments are destroyed at 
increasing excitation energy and their neutrons are combined into the IMFs, 
since the number of free neutrons is still small at this stage. 
Simultaneously the N/Z ratio of the heaviest fragments decreases slightly. 
At very high excitation energy ($E_s^* > 8$ MeV/nucleon) the N/Z ratio of 
IMFs does not rise anymore but drops too because no heavier 
fragments are left and the number of free neutrons increases fastly. 

We note that this isospin evolution takes place in the energy range which is 
usually associated with a liquid-gas type phase transition in finite nuclei 
\cite{PR95,dagostino99}. Our calculations show how the isospin 
fractionation phenomenon \cite{mueller} associated with this phase 
transition actually shows up in finite nuclear systems. This effect of 
a "rise-and-fall" of the IMF's neutron content (see also Fig.~4 in 
\cite{milazzo}) is 
sensitive to the details of the description of hot primary fragments (e.g. 
binding energy, shell effects) as well as to the size and isospin 
content of the 
thermal source. Therefore, it is a good tool for probing the freeze-out 
conditions. Moreover, this effect can help to reveal the difference between 
dynamical and statistical mechanisms of multifragmentation. As was shown in 
many papers (e.g. \cite{nebauer99,larionov99}), both approaches give 
nearly similar results concerning the IMF yields. However, the dynamical 
models favor the opposite isospin effect by decreasing the N/Z ratio 
of larger fragments and thus making them more symmetric 
\cite{larionov99,ditoro}. In such a case the dynamics provides a decreasing 
neutron content of IMFs for larger excitation energies of sources (see e.g. 
\cite{ditoro}) contrary to the experimental observation \cite{milazzo}. 

\begin{figure}
\includegraphics[width=16cm]{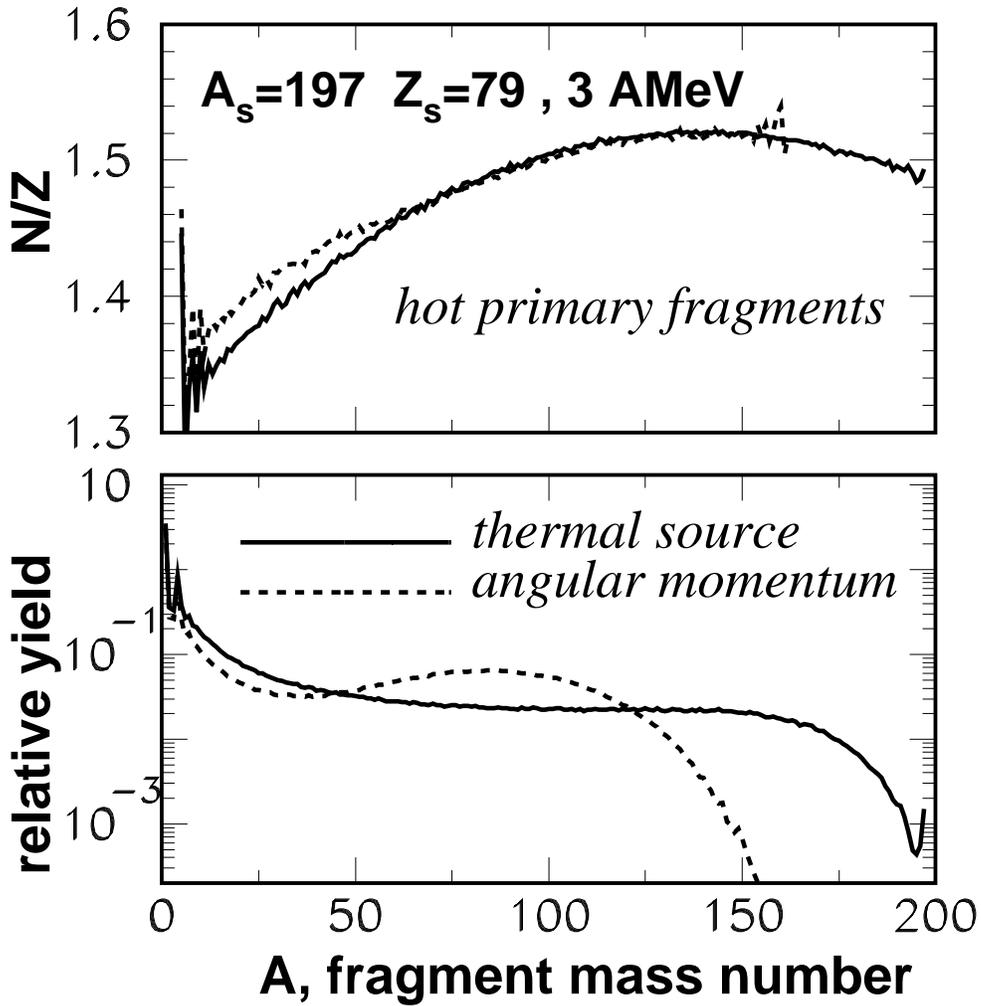}
\caption{The neutron-to-proton ratio N/Z and relative yield of 
hot primary fragments produced after break-up of Au nucleus.
Solid lines: Markov chain SMM calculations for a thermal source with 
excitation energy 3 MeV/nucleon, dashed lines: the same source with angular 
momentum 150$\hbar$.}
\end{figure}

It is instructive to see how an angular momentum influences the isospin of 
fragments emitted from a single source. In this calculation we assume that 
the total excitation energy and angular momentum are fixed and that the 
sharing between rotational and thermal energies is governed by the 
statistical weights of different configurations. In Fig.~3 we show yields and 
N/Z ratios of hot primary fragments produced in the freeze-out volume 
after break-up of Au source at $\rho_s$=$\rho_0$/6 and 
$E_s^{*}$=3 MeV per nucleon. It is seen that an angular momentum favors 
fission-like fragment partitions with two large equal-size fragments (see also 
Refs.\cite{gross97,bot_gro95}). That is different from a normal fragmentation 
pattern dominated by partitions with different-size fragments. An angular 
momentum leads to an increasing N/Z ratio of IMFs as well. 
The latter effect is important and has a simple qualitative explanation. 
An angular momentum favors the emission of IMFs with larger mass numbers, 
providing a larger moment of inertia, in oder to minimize the rotational 
energy and maximize the entropy. On the other side the Coulomb interaction 
prevents the emission of IMFs with large charges. As a result of interplay of 
these two factors we obtain the increasing N/Z ratio. 

In peripheral nucleus--nucleus collisions at the projectile energies of 
10--100 MeV/nucleon the break-up of highly excited projectiles-like nuclei is 
fast (the characteristic time is around 100 fm/c) and happens in 
the vicinity of a target-like residue. The Coulomb field of the target 
residue influences the fragmentation of the projectile source by increasing 
the charge asymmetry of produced fragments and leading to a non-isotropic 
fragment emission: small fragments are preferably emitted to the side of 
the target \cite{botvina99}. Within the Markov chain SMM one can study how 
this effect influences the isotope composition of fragments.

Calculations were performed for the same Au source as in Fig.~3. 
The source was placed at a fixed distance (20 fm) from another Au source 
simulating the target residue. This 
distance was obtained under assumption that the break-up happens at 
$\sim$100 fm/c after a peripheral collision of 35 $A\cdot$MeV Au projectile 
with Au target. 
In reality the break-up may happen at various distances, excitation 
energies and angular momenta. In statistical approach we can take 
into account the distribution of sources in distances and other 
characteristics by considering an {\it ensemble} of the 
sources. Parameters of this ensemble can be found by global comparison with 
the experiment \cite{PR95,deses98}. However, the approximation of a 
fixed distance is sufficient for the qualitative identification of a new 
statistical effect. 

\begin{figure}
\includegraphics[width=16cm]{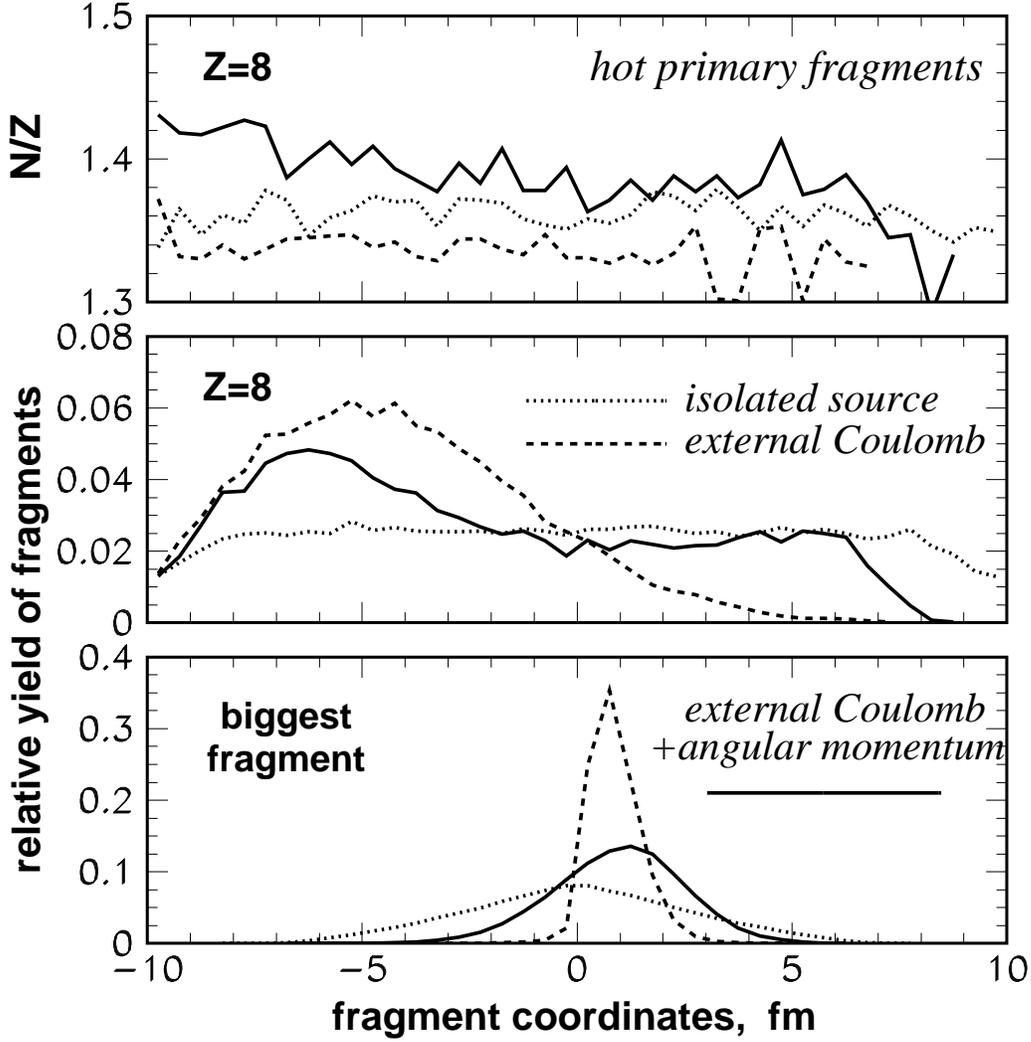}
\caption{Freeze-out coordinate distributions of neutron-to-proton 
ratio N/Z of primary fragments with Z=8 (top panel) and relative yields of 
the primary Z=8 and biggest fragments (middle and bottom panels) produced 
at break-up of Au source at excitation energy of 3 MeV/nucleon. 
The second Au nucleus is placed at -20 {\em fm} from the center of the 
freeze-out volume. 
Dotted lines: the isolated Au source, dashed lines: Coulomb interaction  
with the second Au nucleus is included, solid lines: angular momentum of 
150 $\hbar$ is included as well.}
\end{figure}

In Fig.~4 we show the spatial distributions of yields and N/Z ratios 
of hot primary IMFs 
with $Z$=8 and the biggest fragments in the freeze-out volume along the 
axis connecting the projectile and target sources. It is seen that in the case 
of a single isolated source all distributions are symmetric with respect to 
the center of mass of the source. In the case of an 
external Coulomb field induced by the target source, the IMFs are mainly 
located at the target side while the biggest fragments are shifted to the 
opposite direction. Such positioning of fragments minimizes the Coulomb 
energy of the target-projectile system. However, the external Coulomb field 
alone cannot affect essentially the fragment isospin distribution. 
In case of an additional angular momentum the N/Z ratio of the IMFs increases 
considerably and becomes larger for the IMFs which are closer to the target. 
The reason is again an interplay between the Coulomb and rotational energy. 
To maximize the moment of inertia the system needs more heavy IMFs, 
while to minimize the Coulomb energy, depending on the fragment distance to 
the target, the IMFs should have smaller charges.

Our calculations show that this asymmetry of the IMF isospin distribution 
survives after secondary 
deexcitation of hot fragments. The following Coulomb propagation pushes 
the IMFs in the direction of the target providing predominant population 
of the midrapidity kinematic region by neutron-rich fragments. The Coulomb 
repulsion from the sources may be not sufficient to accelerate fragments up 
to the energy corresponding to the center of mass velocity of the colliding 
nuclei. Nevertheless, the 
IMFs can populate a considerable part of the 
midrapidity region \cite{botvina99}. Within the statistical 
picture a slight radial flow can supply the IMFs with additional velocities to 
fill the central part of the midrapidity zone. Such mechanism of the 
neutron-rich IMF production is consistent with 
the experimental data \cite{dempsey,larochelle}.

In conclusion, a new Markov chain version of the SMM was applied for 
analysis of the isotope composition of fragments and its evolution. 
The increasing N/Z ratio of IMFs with excitation energy can be considered 
as a manifestation of the isospin fractionation associated with the phase 
transition in finite nuclei. 
It was also shown that in peripheral nucleus--nucleus collisions the 
characteristics of statistically produced fragments depend on Coulomb 
interaction between the target- and projectile-like sources and on angular 
momentum transferred to the sources. In particular, this leads to spatial 
asymmetry of both the fragment emission and their isotope composition 
respective to the source center of mass. Previously the 
symmetry violation was considered as a sign of the dynamical emission from 
the neck region. However, we have demonstrated that there exists an 
alternative statistical explanation: 
the symmetry of the phase space population is broken by the interaction 
between two sources. Theoretically such a process gives example of a new 
kind of statistical emission influenced by an inhomogeneous external 
long--range field \cite{botvina99}. 

A.S.B. thanks GSI for financial support and warm hospitality. I.N.M. 
thanks the Humboldt Foundation for the financial support.


\begin{thebibliography}{99}
\bibitem{hirschegg} See for instance:
"Multifragmentation", Proceedings of the International
Workshop 27 on Gross Properties of Nuclei and Nuclear Excitations,
Hirschegg, Austria, January 17--23, 1999. GSI, Darmstadt, 1999.
\bibitem{PR95} J.P.Bondorf, A.S.Botvina, A.S.Iljinov, I.N.Mishustin and 
K.Sneppen. {\em Phys. Rep.} {\bf 257}, 133 (1995).
\bibitem{gross97} D.H.E.Gross, {\em Phys.Rep.} {\bf 279}, 119 (1997).
\bibitem{botvina95} A.S.Botvina {\it et al.}, {\em Nucl.Phys.} {\bf A584} 
(1995) 737.
\bibitem{dagostino99} M.D'Agostino {\it et al.}, {\em Nucl. Phys.} {\bf A650}, 
329 (1999).
\bibitem{dagostino96} M.D'Agostino {\it et al.}, {\em Phys. Lett.} {\bf B371}, 
175 (1996).
\bibitem{williams} C.Williams {\it et al.} {\em Phys. Rev.} {\bf C55}, 
R2132 (1997).
\bibitem{botvina87} A.S.Botvina {\it et al.}, {\em Nucl.Phys.} {\bf A475} 
(1987) 663.
\bibitem{mc_pre} A.S.Botvina, A.D.Jackson and I.N.Mishustin,
{\em Phys. Rev.} {\bf E62}, R64 (2000). 
\bibitem{mc_smm} A.S.Botvina, I.N.Mishustin {\it et al.}, in preparation.
\bibitem{markov} K.L. Chung, Markov processes with stationary transition 
probability. Springer, Heidelberg, 1967.
\bibitem{bot_gro95} A.S.Botvina and D.H.E.Gross, {\em Nucl. Phys.} {\bf A592}, 
257 (1995).
\bibitem{viola} P.Marmier and E.Sheldon, Physics of Nuclei and Particles. 
Academic Press, N.Y., 1969 
\bibitem{milazzo} P.M.Milazzo, A.S.Botvina, G.Vannini et al., {\em Phys. Rev.} 
{\bf C62}, 041602(R) (2000). 
\bibitem{mueller} H.M\"uller and B.D.Serot, {\em Phys. Rev.} {\bf C52}, 2072 (1995)
\bibitem{nebauer99} R.Nebauer, J.Aichelin {\it et al.}, {\em Nucl. Phys.} 
{\bf A658}, 67 (1999).
\bibitem{larionov99} A.B.Larionov, A.S.Botvina, M.Colonna and M.Di~Toro, 
{\em Nucl. Phys.} {\bf A658}, 375 (1999).
\bibitem{ditoro} M.Di~Toro {\it et al.}, nucl-th/0009079, 2000. 
\bibitem{botvina99} A.S.Botvina, M.Bruno, M.D'Agostino and D.H.E.Gross, 
 {\em Phys. Rev.} {\bf C59}, 3444 (1999).
\bibitem{deses98} P.Desesquelles {\it et al.}, {\em Nucl. Phys.} {\bf A633}, 
547 (1998).
\bibitem{dempsey} J.F.Dempsey {\it et al.}, {\em Phys. Rev.} {\bf C54}, 1710 (1996).
\bibitem{larochelle} Y.Larochelle {\it et al.}, {\em Phys. Rev.} {\bf C62}, 
051602(R) (2000).
\end{thebibliography}
\end{document}